# AN OVERVIEW OF THE SPALLATION NEUTRON SOURCE PROJECT

Robert L. Kustom, SNS/ORNL, Oak Ridge, TN

*Abstract*

The Spallation Neutron Source (SNS) is being designed, constructed, installed, and commissioned by the staff of six national laboratories, Argonne National Laboratory, Brookhaven National Laboratory, Jefferson National Accelerator Laboratory, Lawrence Berkeley National Laboratory, Los Alamos National Laboratory, and Oak Ridge National Laboratory. The accelerator systems are designed to deliver a 695 ns proton-pulse onto a mercury target at a 60-Hz repetition rate and an average power of 2-MW. Neutron moderators that will convert the spallation neutrons into slow neutrons for material science research will surround the target. Eighteen neutron beam lines will be available for users, although initially, only 10 instruments are planned. The Front-End Systems are designed to generate a 52 mA, H- beam of minipulses, 68% beam on, 32% beam off, every 945 ns, at 2.5 MeV for 1 ms, 60 times a second. The Front-End systems include a RF driven, volume-production ion source, beam chopping system, RFQ, and beam transport. The linac consists of a drift tube linac up to 86.8 MeV, a coupled-cell linac to 185.7 MeV, and a superconducting RF linac to the nominal energy of 1 GeV. The design of the superconducting section includes 11 cryomodules with three, 0.61-beta cavities per cryomodule and 15 cryomodules with four, 0.81-beta cavities per cryomodule, with space to install six more 0.81-beta cryomodules. The accumulator ring is designed for charge exchange injection at full energy and will reach 2.08x10E+14 protons/pulse at 2-MW operation. The goal is to reduce uncontrolled beam losses to less than 1x10E-4. A detailed overview of the accelerator systems and progress at the various laboratories will be presented.

## 1 INTRODUCTION

The Spallation Neutron Source (SNS) facility under construction at Oak Ridge National Laboratory is designed to generate pulses of neutrons at intensities well beyond any of the world's existing spallation neutron sources. The accelerator systems are designed to deliver a 695 ns proton-pulse onto a liquid mercury target at a 60-Hz repetition rate with an average proton beam power of 2-MW. The target station will have 18 shutters that ultimately will be able to support 24 neutron instruments. An initial complement of ten instruments is planned at the start of operation in 2006. A site master plan is shown in Figure 1.

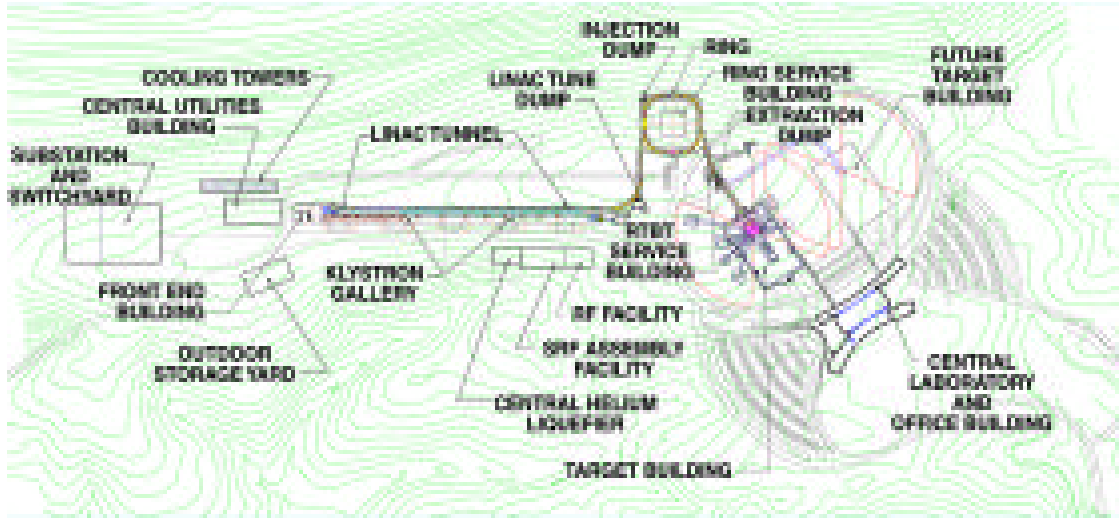

Figure 1: Site Master Plan.

The SNS is being designed and built as a partnership of six DOE national laboratories: Lawrence Berkeley (LBL) in California, Los Alamos (LANL) in New Mexico, Argonne (ANL) in Illinois, Oak Ridge (ORNL) in Tennessee, Brookhaven (BNL) in New York, and Thomas Jefferson (JLAB) in Virginia. The Front-End Systems (FES) are the responsibility of LBL. The drift tube linac (DTL), coupled-cell linac (CCL), and warm parts of the linac, including the end-to-end physics design and RF system design are the responsibility of LANL. The superconducting RF cavities, cryomodules, and cryogenic equipment are the responsibility of JLAB. The accumulator ring and high-energy transport lines between the linac and the ring (HEBT) and the ring and the target (RTBT) are the responsibility of BNL. The target station and conventional facilities are the responsibility of ORNL. The neutron instruments are the responsibility of ANL. Project integration, direction, and planning for operation are the responsibilities of the SNS office at ORNL This article describes the combined effort on the part of staff at these laboratories.

Considerably more detail is provided in a number of excellent papers being presented at this conference.

A summary of key design parameters for the SNS facility is presented in Table 1.

Table 1. Summary of key design parameters for the SNS Facility

| | |
|---|---|
| Proton beam power on target, Mw | 2 |
| Average proton beam current on target, mA | 2 |
| Pulse repetition rate, Hz | 60 |
| Chopper beam on duty factor, % | 68 |
| Front-end and linac length, m | 335 |
| DTL output energy, MeV | 87 |
| DTL frequency, MHz | 402.5 |
| CCL output energy, MeV | 185 |
| Number of SRF cavities | 92 |
| Linac output energy, GeV | ≈1 |
| CCL and SRF frequency, MHz | 805 |
| Linac beam duty factor, % | 6 |
| High Energy Beam Transport (HEBT) length, m | 170 |
| Accumulator ring (AR) circumference, m | 248 |
| Ring orbit revolution time, ns | 945 |
| Number of turns injected into AR during fill | 1060 |
| AR fill time, ms | 1 |
| Gap in AR circulating beam for extraction, ns | 250 |
| Length from AR to production target (RTBT), m | 150 |
| Peak number of accumulated protons per fill | 2.08E+14 |
| Proton pulse width on target, ns | 695 |
| Target material | Hg |
| Number of neutron beam shutters | 18 |
| Initial number of instruments | 10 |
| Number of instruments for complete suite | 24 |

## 2 TECHNICAL DESIGN OF THE ACCELRATOR SYSTEMS

### *2.1 Front-End Systems*

The Front-end Systems (FES) are designed to generate an H- beam of mini-pulses with 68% on time, every 945 nanoseconds for a period of 1 millisecond at a 60 Hz repetition rate. The FES include a RF driven, volume-production ion source, beam chopping system, RFQ, a low-energy beam transport (LEBT) system, and a medium-energy beam transport (MEBT) system. The FES must deliver 52 ma at 2.5 MeV at the input to the drift tube linac. The key FES parameters are listed in Table 2[1].

The H- ion source utilizes a 2-MHz, RF driven discharge to generate the plasma. The plasma is confined by a multi-cusp magnet configuration. A magnetic dipole filter separates the main plasma from the region where low-energy electrons generate the negative ions. A heated cesium collar surrounds the production chamber. Electrons are removed from the ion beam by a deflecting field from a dipole magnet arrangement in the outlet plate of the plasma generator. The ion source is tilted with respect to the LEBT to compensate for the effect the electron-clearing field has on the ion beam.

The LEBT structure is based on an earlier design [2] that proved the viability of purely electrostatic matching. There are two einzel-lens in the LEBT. The second is split into quadrants that can be biased with D.C. and pulsed voltages to provide angular steering and pre-chopping. Chopping voltages of +- 2.5 kV and 300 ns are rotated around the quadrants. Corrections in transverse beam displacement are achieved by moving the ion source and LEBT with respect to the RFQ [3]. A schematic of the ion source and LEBT that will be used for startup of the facility is shown in Fig. 1. Its performance goal is 35 ma, and it will be a significant step towards developing the full 65 mA estimated for 2 MW operation.

Table 2. FES Key Performance Parameters

| | |
|---|---|
| Ion Species | H- |
| Output Energy, MeV | 2.5 |
| H- current @ MEBT output, mA | 52 |
| Nominal H- current @ ion-source output, mA | 65 |
| Output normalized transverse rms emittance, π mm mrad | 0.27 |
| Output normalized longitudinal rms emittance, π MeV deg | 0.13 |
| Macro pulse length, ms | 1 |
| Duty factor, % | 6 |
| Repetition rate, Hz | 60 |
| Chopper rise & fall time, ns | 10 |
| Beam off/beam on current ratio | 10E-4 |

The RFQ will accelerate beam from 65 keV to 2.5 MeV with an expected transmission efficiency of better than 80%. It is built in four modules using composite structures with a GlidCop shell and four oxygen-free-copper vanes. The length of the RFQ is 3.72 m.

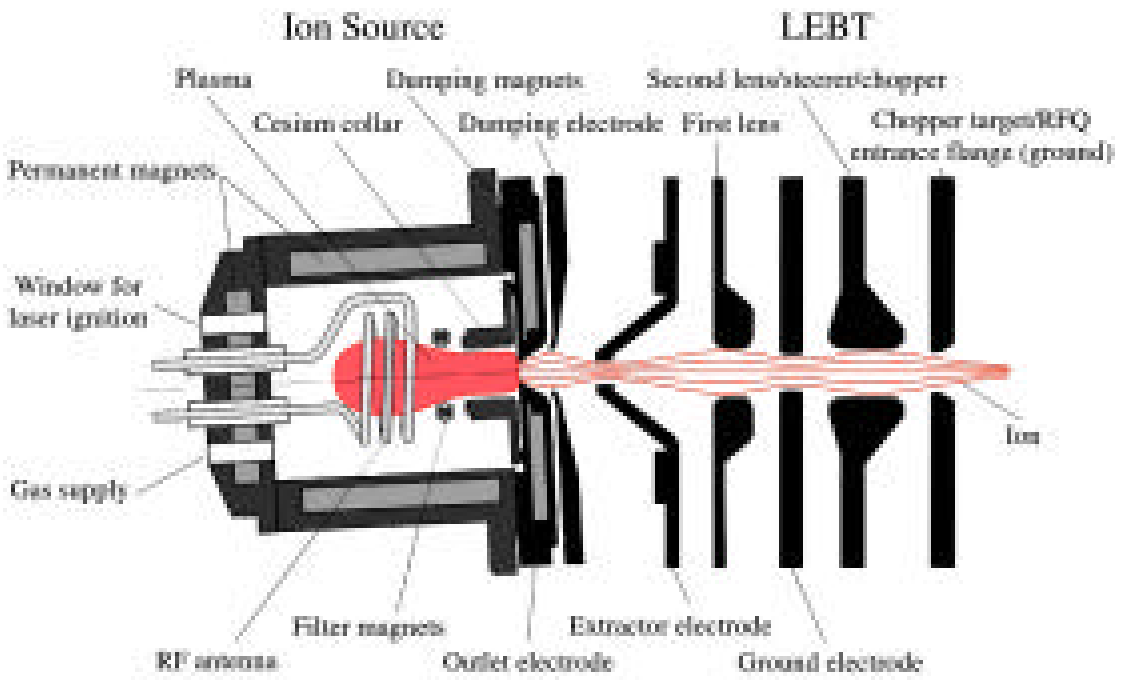

Figure 2: Schematic of the startup ion.

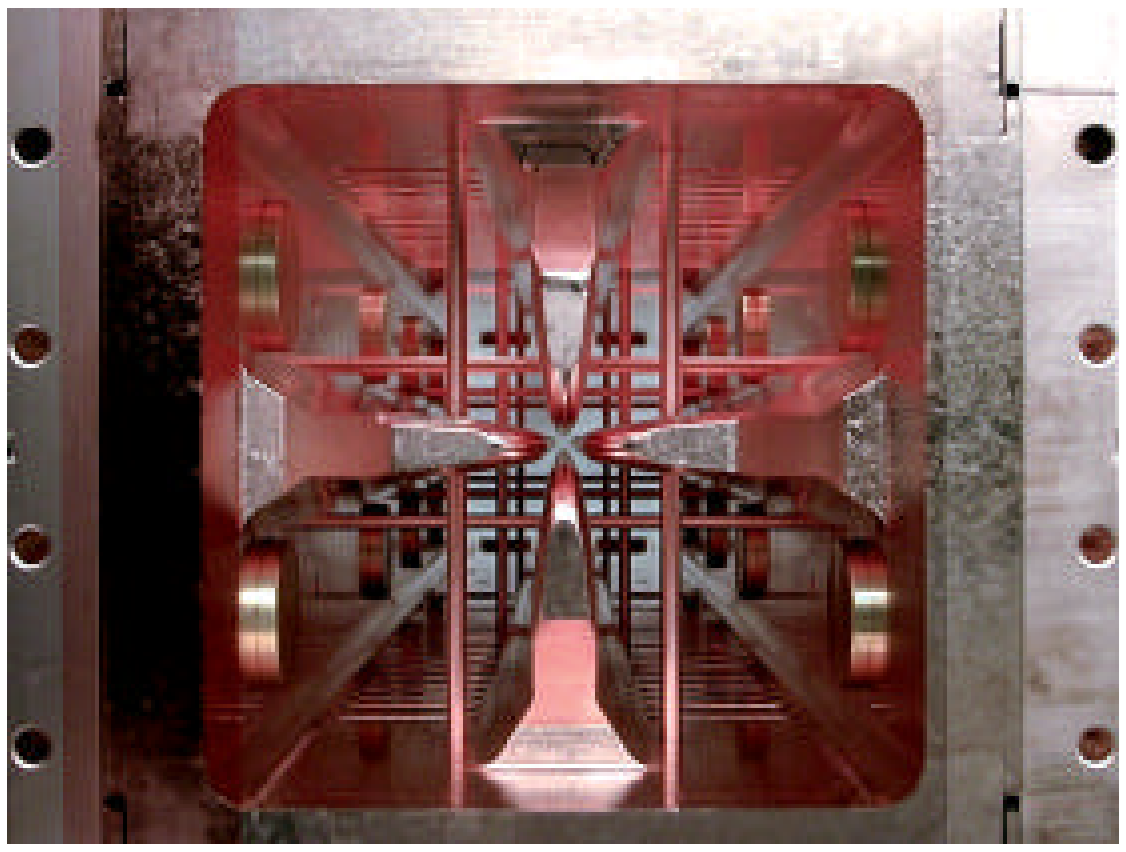

Figure 3: End-on view of the assembled RFQ source for the SNS module.

The design frequency is 402.5 MHz. Peak surface fields reach 1.85 Kirkpatrick and require 800 kW during the pulse. The output of RFQ is directed into the MEBT [4]. Matching from the RFQ to DTL is performed in the MEBT. Final chopping of the bunches is also performed in the MEBT.

### *2.2 Linac Systems*

The linac consists of a drift tube linac up to 86.8 MeV, a coupled-cell linac (CCL) up to 185.7 MeV, and a superconducting linac up to a nominal energy of 1 GeV. The superconducting linac is divided into a medium-beta cavities and high-beta cavity sections [5]. The medium-beta cavity is designed for a geometric $\beta$ of 0.61, and the high-beta cavity is designed for a geometric $\beta$ of 0.81. The nominal transition energy between the medium and high beta sections is 378.8 MeV.

The DTL consists of six separate tanks each driven by a 402.5 MHz, 2.5 Mw klystron. The focusing lattice is FFODDO with a six $\beta\lambda$ period. The focusing magnets are permanent magnet quadrupoles with constant GL of 3.7 kG and a bore radius of 1.25 cm. There are one-$\beta\lambda$ inter-tank gaps for diagnostics. Empty drift tubes contain BPMs and steering dipoles. There are 144 quadrupoles and 216 drift tubes in the DTL. The energy gain per real estate meter is 2.3 MeV/m in the DTL. Key parameters for the DTL are listed in Table 3.

The CCL operates at 805 MHz. There are eight accelerating cells brazed together to form a segment. Six segments are mounted and powered together as a single module using 2.5-$\beta\lambda$ coupling cells, one of which is powered. A 3-D schematic of Module 1 is shown in Fig. 4 and a cutaway view of the segments and the powered coupler is shown in Fig 5.

Table 3: DTL Parameters

| Tank # | Final Energy (MeV) | Power (Mw) | Length (m) | # of cells |
|---|---|---|---|---|
| 1 | 7.46 | 0.52 | 4.15 | 60 |
| 2 | 22.83 | 1.6 | 6.13 | 48 |
| 3 | 39.78 | 1.93 | 6.48 | 34 |
| 4 | 56.57 | 1.93 | 6.62 | 28 |
| 5 | 72.49 | 1.87 | 6.54 | 24 |
| 6 | 86.82 | 1.88 | 6.61 | 22 |

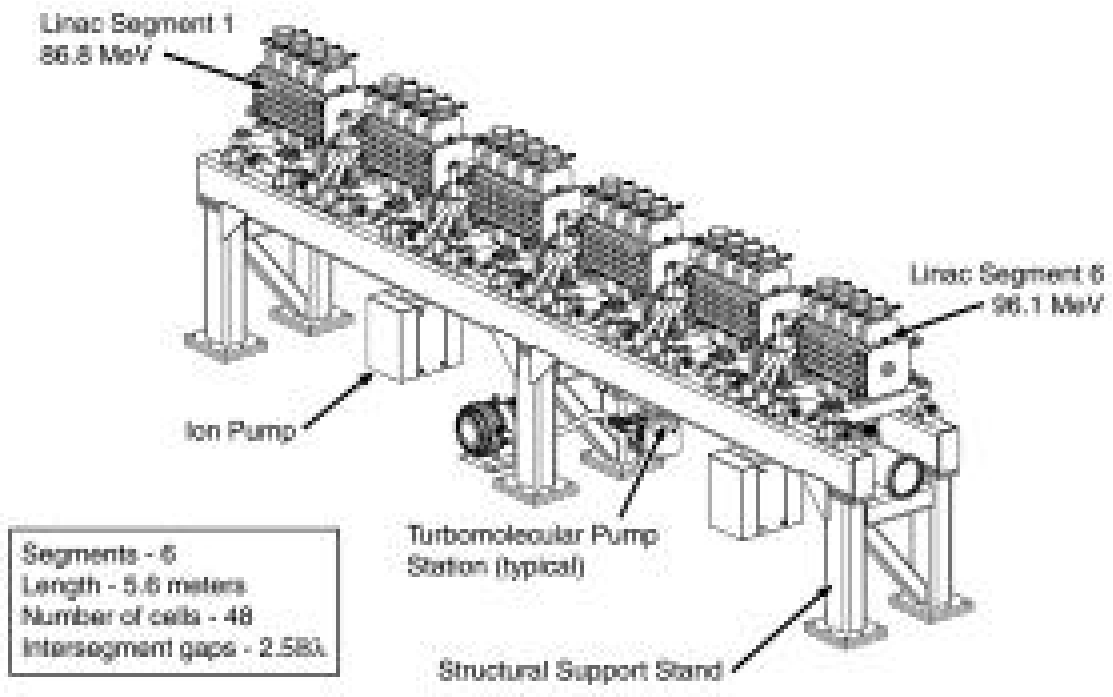

Figure 4: 3-D Schematic of CCL module 1.

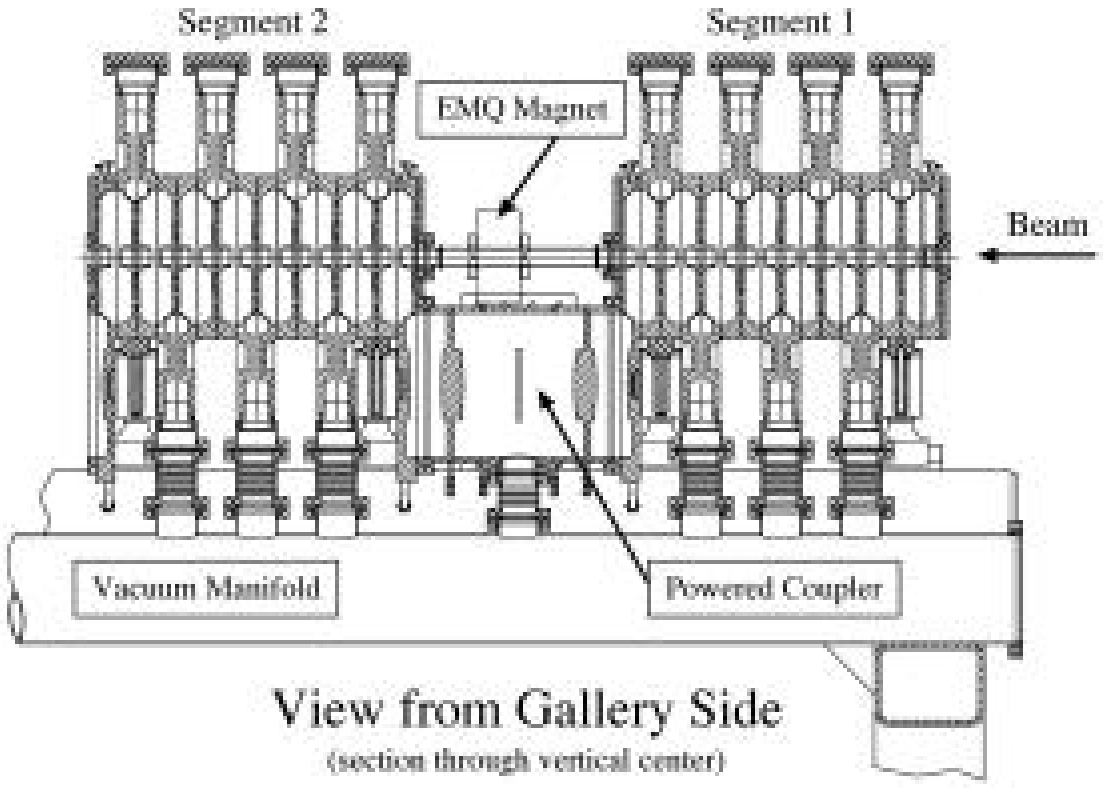

Figure 5: Cutaway view of CCL through segments 1 & 2 and the powered coupler.

There are a total of eight modules. Four, 5-megawatt klystrons drive the CCL. Each klystron drives two modules. The peak power is 11.4 Mw and the maximum accelerating field on axis is 3.37 MV/m (EoT). The energy gain per real estate meter is 1.7 MeV/m. The transverse focusing system is a FODO lattice in the CCL. The bore radius goes from 1.5 cm to 2.0 cm. The total length of the CCL is 55.12 m.

The high-energy end of the linac, above 185.7 MeV, uses superconducting cavities. The design is based on a conceptual design study completed by scientists from many institutions and lead by Yanglai Cho [6].

Two different superconducting cavity designs are used in the SNS linac, one with a geometric β of 0.61, defined as the medium-β cavity, and the other a geometric β of 0.81, defined as the high-β cavity. There are six cells per cavity in the medium and high-β sections. More than six cells per cavity results in excessive phase slip for a particular beta and fewer than six cells per cavity results in inefficient use of real estate and higher cost due to increased parts count. The cavities will be fabricated using 4 mm-thick Nb with stiffening or reinforcement plates. The initial design assumes a peak field of 27.5 MV/m, +-2.5 MV/m, however, with conditioning and future processing, higher gradients are expected. The design value for Qo is 5x10E+9, and the loaded Q design value is 5x10E+5. The effective accelerating gradients are 10.5 MV/m in the 0.61-β section and 12.8 MV/m in the 0.81-β section. The design values for Lorentz detuning, referenced to the geometric accelerating field, are 2.9 Hz/(MV/m)^2 in the medium-β cavities and 1.2 Mz/(MV/M)^2 in the high-β cavities. The 6-σ design value for microphonics is +-100 Hz. Cold tuning will allow the cavities to be taken off resonance by 100 kHz. Each cavity is driven by a single, 550 kW klystron operating at 805 MHz [7].

There are three cavities per cryomodule in the medium-β section, and a total of eleven medium-β cryomodules in the linac. There are four cavities per cryomodule in the high-β section of the linac. Initially, fifteen high-β cryomodules will be installed. There are, however,

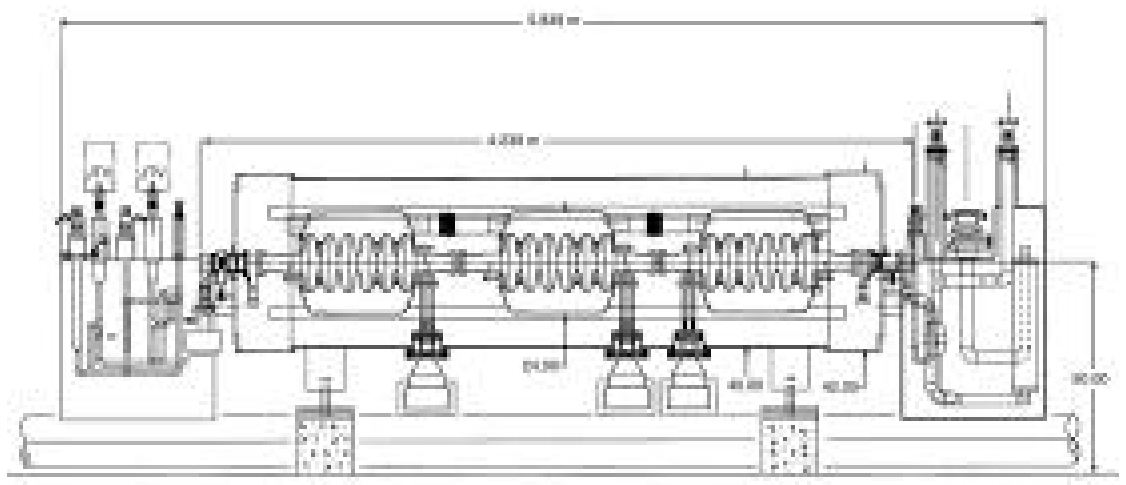

Figure 6: Schematic view of a medium-β cryomodule and superconducting cavities.

additional straight-section spaces to install as many as twenty-one high-β cryomodules in the future. A schematic sectional view of the medium-β cryomodule through the superconducting cavities is shown in Fig. 6

A summary of the key superconducting linac dimensions is listed in Table 4 and key cryogenic parameters are listed in Table 5.

Table 4. Key superconducting RF cavity dimensions

| | |
|---|---|
| Nb thickness, mm | 4.0 |
| Minimum bore radius, medium-β, cm | 4.3 |
| Cryomodule length, medium- β, m | 4.239 |
| Cryomodule length, high-β, m | 6.291 |
| # of medium-β cryomodules | 11 |
| # of high-β cryomodules (initial) | 15 |
| Warm space between cryomodule cells, m | 1.6 |
| Total length of SRF linac with extra 6 cryomodules, m | 235.92 |

Table 5. Cryogenic requirements for superconducting linac

| | |
|---|---|
| Operating temperature, K | 2.1 |
| Primary circuit static load, w | 785 |
| Primary circuit dynamic load, w | 500 |
| Primary circuit capacity, w | 2500 |
| Secondary circuit temperature, K | 5.0 |
| Secondary circuit static load, g/s | 5 |
| Secondary circuit dynamic load, g/s | 2.5 |
| Shield circuit temperature, K | 35-55 |
| Shield circuit load, w | 5530 |
| Shield circuit capacity, w | 8300 |

*2.3 Accumulator Ring*

The accumulator ring for SNS is a FODO arc with doublet straight sections [8]. This lattice has four-fold symmetry with zero dispersion in the straight sections. A plan view of the ring and transport lines is shown in Fig. 7. The ring circumference is 248 m. The zero dispersion regions include two-6.85 meter sections and one long 12.5 meter section. Each of the four straight sections has a dedicated function. The injection straight includes the injection septum magnet, eight bump magnets for horizontal and vertical injection painting, the stripper foil, and dump septum. The collimator section includes moveable scattering foils and three fixed collimators. The extraction section includes fourteen full-aperture-ferrite, extraction kicker magnets and a Lambertson extraction septum magnet. The rise time of the extraction kickers is 200 ns. The RF section has three first-harmonic cavities operating at 1.058 MHz and a second harmonic cavity. The total voltage generated at the first harmonic is 40 kV and at the second harmonic is 20 kV.

At 2 MW operation, 2.08x10E14 protons are accumulated in a 650-700 ns bunch in 1060 turns. The injection process is direct charge exchange using a painting scheme to achieve uniform transverse charge distribution and a second harmonic RF system to spread the beam more uniformly in the longitudinal plane. The expected fractional space-charge tune-shift is 0.14. The goal for gap cleanliness is 10E-4 beam-in-gap/total beam.

Achieving low uncontrolled beam loss, less than 10E-4, is a key element of the accumulator ring design. The design of the injection process, collimation scheme, RF system design, emittance and acceptance ratio, and extraction system are all designed to achieve this low level of beam loss.

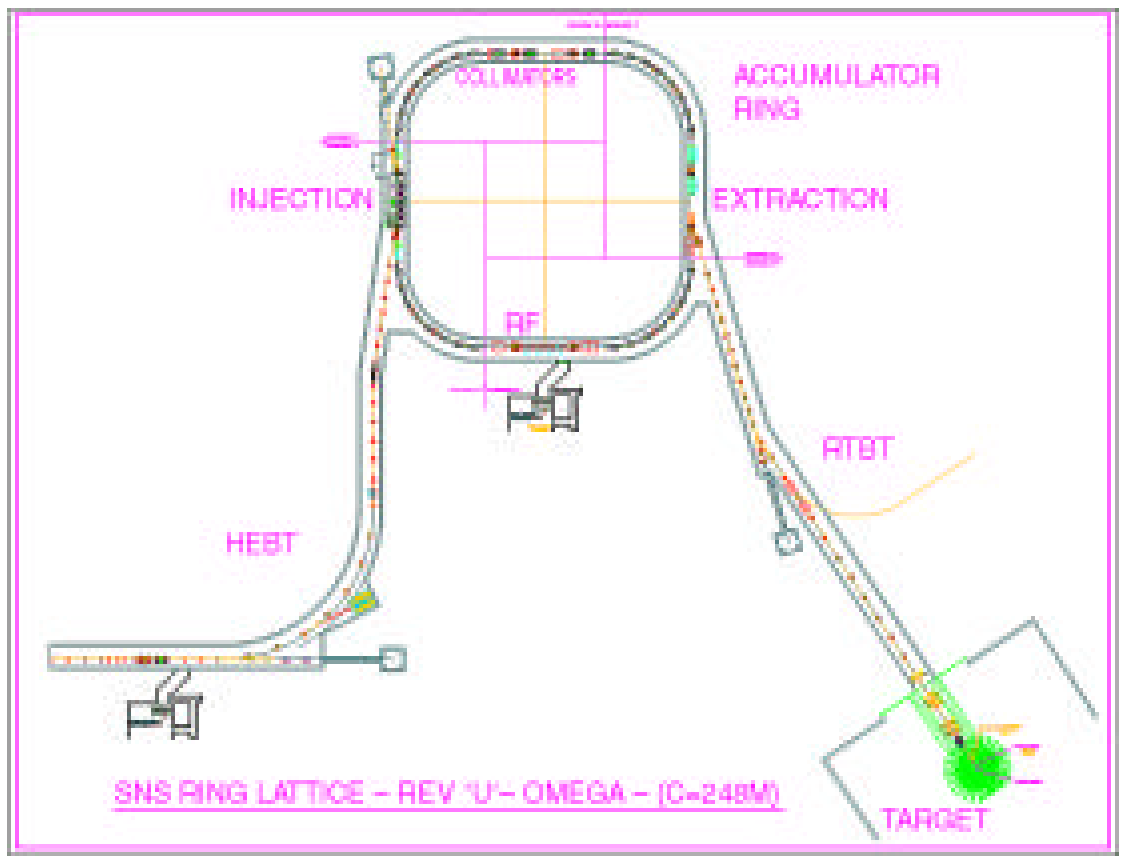

Figure 7: Plan view of the SNS accumulator ring.

## 3 PROJECT STATUS

Major construction has started on the conventional facilities and the technical components. Excavation on the site is currently about 40% complete. Much of the building and utility detailed design has started and major civil procurements, such as bulk concrete and structural steel, are well along in the procurement cycle. The start-up ion source and all electrostatic LEBT have been successfully operated at LBL at 42 mA, greater than the initial 35 mA needed for the start of commissioning. The first of the RFQ modules has been fabricated and tested at full field and pulse length. A cold model of the DTL is in fabrication and a cold model of the CCL has been successfully tested at LANL. A significant number of major linac procurements, such as the 402.5 MHz, 5 MW klystrons, 402.5 MHz circulators, and transmitters for klystron control, have been awarded. The copper model for the 0.61-β single cell has been brazed and is being tested at JLAB. Six-cell Nb cavities are being fabricated. The procurement of Nb for construction of all the cavities and much of the hardware for the cryogenic facility and cryomodule production has been awarded. Procurement of ferrite for the ring RF systems has been awarded and sample is being tested. Ring dipole, quadrupole, and corrector magnets have tested, and procurement of these magnets has started at BNL.

In summary, major construction has started and the project expects to meet the goal of first beam injected into the accumulator ring by July 2004, and first beam on target by January 2005.